\documentclass{IEEEtran}
\usepackage{cite}
\usepackage{amsmath,amssymb,amsfonts}
\usepackage{algorithmic}
\usepackage{graphicx}
\usepackage{textcomp}

\usepackage{url}
\usepackage{graphicx}
\usepackage{xcolor}
\usepackage{placeins}
\usepackage{float}
\usepackage{tabularx,colortbl}
\usepackage{ifthen}
\usepackage{mathtools} 
\usepackage{array}
\usepackage{booktabs}
\newcolumntype{P}[1]{>{\centering\arraybackslash}p{#1}}

\usepackage[Symbol]{upgreek}
\usepackage[greek,english]{babel}
\renewcommand{\pi}{\textrm{\greektext p}}

\newcommand{\RE}[1]{{\rm Re}\{#1\}}

\usepackage{hyperref}
\hypersetup{
	colorlinks=true, 
	linkcolor=blue!70!black, 
	citecolor=red!70!black, 
	filecolor=blue!70!black, 
	urlcolor=blue!70!black, 
}

\newcommand{\Z}{{\bf Z}}
\newcommand{\M}{{\bf M}}
\newcommand{\Q}{{\bf Q}}

\newcommand{\vv}{{\bf v}}
\newcommand{\iv}{{\bf i}}
\newcommand{\wv}{{\bf w}}
\newcommand{\rv}{{\bf r}}
\newcommand{\Ev}{\overline{E}}
\newcommand{\GAUact}{G^\mathrm{a}_\mathrm{AU}} 
\newcommand{\GBUact}{G^\mathrm{a}_\mathrm{BU}} 
\newcommand{\GAU}{G_\mathrm{AU}} 
\newcommand{\GBU}{G_\mathrm{BU}} 

\providecommand{\herm}{\mathrm{H}}
\providecommand{\trans}{\mathrm{T}}
\providecommand{\J}{\ensuremath{\mathrm{j}}}
\providecommand{\T}[1]{\mathrm{#1}}

\DeclareRobustCommand{\cl}[1]{{#1}}
\DeclareRobustCommand{\clrev}[1]{{#1}}

\begin{document}

\title{\cl{Generalized Friis Transmission Formula Using Active Antenna Available Power and Unnamed Power Gain}}

\author{Karl F.\ Warnick,~\IEEEmembership{Fellow, IEEE}, Frederic Broyde,~\IEEEmembership{Senior Member, IEEE}, Lukas Jelinek,\\Miloslav Capek,~\IEEEmembership{Senior Member, IEEE}, and Evelyne Clavelier,~\IEEEmembership{Senior Member, IEEE}

\thanks{This material is based upon work supported by the National Science Foundation under Grant No.\ 163664 and by the Czech Science Foundation under project~\mbox{No.~21-19025M}.} 
\thanks{Karl Warnick is with the Department of Electical and Computer Engineering, Brigham Young University, Provo, UT USA (e-mail: warnick@byu.edu).}
\thanks{Fr\'{e}d\'{e}ric~Broyd\'{e} is with Eurexcem, 12 chemin des Hauts de Clairefontaine, 78580 Maule, France (e-mail: fredbroyde@eurexcem.fr).}
\thanks{L. Jelinek and M. Capek are with Czech Technical University in Prague, Czech Republic (e-mails: lukas.jelinek@fel.cvut.cz, miloslav.capek@fel.cvut.cz).}
\thanks{Evelyne~Clavelier is with Excem, 12 chemin des Hauts de Clairefontaine, 78580 Maule, France (e-mail: eclavelier@excem.fr).}}

\maketitle

\begin{abstract}
\cl{We use the concept of active antenna available power to derive a generalization of the Friis transmission formula for multiport antenna systems. With beamformer weights chosen such that the array patterns are the same when transmitting and receiving, the active antenna available power at the receiving antenna divided by the input power at the transmitter is symmetric under link direction reversal in the near field as well as the far field. These results generalize the Friis transmission formula to beamformed multiport antenna systems in an arbitrary reciprocal propagation environment.}
\end{abstract}


\section{Introduction}
\cl{
The Friis transmission formula is widely used in the analysis of power transfer between antennas and communication system link budgets. After its introduction in 1946~\cite{friis1946note}, the Friis transmission formula has generalized to handle voltage phasors~\cite{franek2017phasor}, electromagnetic beams \cite{goubau1961guided}, the Fresnel region \cite{kim2013generalised,wang2022group}, and the near-field power transfer for short-range communications \cite{warnick2012optimizing}, aperture antennas \cite{chu1965approximate}, and antenna arrays \cite{kim2019wireless,song2021analysis,wu2022accurate}. 
}

\cl{
Existing generalizations of the Friis formula have several limitations. For multiport antenna systems, most treatments represent the power transfer from elements on one array to another in terms of a complicated sum over individual element responses \cite{song2021analysis,wu2022accurate}. Existing generalizations also lack an explicit symmetry under link direction reversal. A multiple input multiple output (MIMO) communication system, for example, may use different beamformer weights on transmit and receive, so it is unclear how reciprocity applies to the MIMO communication link. The conventional Friis transmission formula applied to reciprocal antennas, on the other hand, clearly gives the same result for the available power at the receiving antenna divided by the input power at the transmitter, with the transmitter and receiver swapped. Ideally, a generalization of the Friis transmission formula to multiport antenna systems would reflect the underlying reciprocity of the antennas in a natural way. Lastly, the Friis transmission formula is inherently a far-field relationship, and while near-field extensions are available, they are unwieldy and in many cases only approximate. 
}

\cl{
For some applications, including MIMO communication systems and astronomical receiving arrays, the Friis formula is too simplistic to be directly useful. In these applications, the S-parameter matrix or impedance matrix formulation is used to analyze multiport antenna systems using network theory \cite{warnick2021network}. The network theory approach to antenna analysis and design has been widely successful, but it lacks the simple insight into system performance that can be found by using the Friis transmission formula to express power transfer as a budget of contributing terms. 
}

\cl{
Our aim is to show how the network theory formulation generalizes the Friis transmission formula to multiport antenna systems. To do this, we need a way to link the network formulation for antenna arrays to received powers and gains. An active array has electronic gain factors in the receiver chains as well as scale factors in the beamforming network. In generalizing the Friis formula to antenna arrays, it is not obvious how these scale factors should influence power transfer between transmitting and receiving arrays. Furthermore, when the arrays are close together, the presence of one array affects the impedances at the other array element ports. Dealing with this complexity is beyond the reach of the traditional Friis formula, which is inherently a far-field relationship and ignores the effect of the presence of the receiving antenna on the input impedance of the transmitting antenna. 
}

\cl{
A rigorous generalization of the Friis transmission formula has been obtained for arrays terminated by arbitrary load networks \cite{ERPEE5,ERPEE7,ERPEE8}. These papers consider the available power at a receiving antenna divided by the input power at the transmitter, which in network analysis is an unnamed power gain (see Sec.\ \ref{sec:unnamed}). The unnamed power gain is governed by a generalized eigenvalue problem for each link direction where the eigenvectors correspond to excitations at the transmitting array. The maximum and minimum values over the transmit excitations of the unnamed power gain do not change when the direction of the link is reversed \cite[Section~XIII]{ERPEE7}, \cite{ERPEE8}, which is a manifestation of reciprocity for multiport antenna systems. 
}



\cl{
In this article, we consider the case of transmitting and receiving arrays with beamforming networks. By combining a network analysis of the system with the concept of active antenna available power, we show that the gains of the transmitting and receiving arrays and the free-space path loss are given by an active antenna unnamed power gain. } The active antenna unnamed power gain can be found from the system mutual impedance matrix and the array excitations. This provides a generalization of the Friis transmission formula for multiport antenna systems in a linear, reciprocal, and time-invariant but otherwise arbitrary propagation environment.

The generalized Friis transmission formula provides new insight into the symmetry of bidirectional links between antenna arrays. When reversing the direction of transmission, there is no necessary connection between the excitations when an array radiates and the combining coefficients used to form a beam on reception, and the generalized Friis formula for multiport antennas gives a different unnamed power gain in each direction. With appropriate constraints on the transmit excitations and receive beamformer coefficients, so that array 1 has the same pattern on transmit and receive and similarly for array 2, we show that the active antenna unnamed power gains for the two directions in a bidirectional multiport communication system become equal. 


\cl{We also show that the generalized Friis transmission formula presented in this paper reduces to the conventional formula in the far-field limit. The embedded element pattern (EEP) formulation \cite{craeye2011review,davidson2022contemporary} can be used to factor the active antenna unnamed power gain factors into the product of antenna gains and the inverse free space path loss, providing a far-field approximation for the transfer function between two antenna arrays.

Finally, we show that the generalized Friis transmission formula presented in this paper, which applies to arrays with beamforming networks and uses the active antenna unnamed power gain, is connected to the earlier generalization proposed in \cite{ERPEE5,ERPEE7,ERPEE8}, which ignores beamforming networks and uses the unnamed power gain.}



\section{Unnamed Power Gain}
\label{sec:unnamed}

\cl{
The Friis transmission formula gives the ratio of the available power $P_\mathrm{A,2}$ at a receiving antenna to the power $P_{\text{in},1}$ accepted by a transmitting antenna:
\begin{equation}\label{eq:friis}
  \frac{P_\mathrm{A,2}}{P_{\text{in},1}} = G_1 G_2 \left( \frac{\lambda}{4 \pi r}\right)^2
\end{equation}
where $G_1$ and $G_2$ are the antenna gains and other quantities are defined as is usual \cite{balanis2016antenna}. This represents a power gain, but one that is not familiar in network analysis. 

In the analysis of two-port networks, five powers are used in defining power gains: the power available from the source, the power accepted by the network, the power available from the network, the power accepted by the load, and the power accepted by the load if it was directly coupled to the source (see Table~\ref{tab}). The resulting operating power gain, transducer power gain, available power gain, and insertion power gain are well-known and widely used.  

The fifth combination, the ratio of available power from the network to power accepted by the network, is rarely used. It has been called maximum available gain \cite{merat}, although this ratio is not the maximum value of the available power gain. For practical purposes, the ratio of available power to power accepted by the network is unnamed and has been referred to as the ``unnamed power gain'' \cite{ERPEE5}. 

This unnamed power gain is precisely what is given by the Friis transmission formula for antenna systems \cite{friis1946note,ERPEE5}. In this paper, we use network theory to generalize the unnamed power gain to multiport antenna systems in the near and far field cases. 
}

\setlength\extrarowheight{5pt}
\begin{table}
    \caption{Gain Quantities Associated With a Two-port Network}\label{tab}
    \begin{tabular}{P{23mm}P{25mm}P{27mm}}
    \toprule
        & Power accepted by load & Available power from network \\\midrule
        Available power from source    & Transducer power gain & Available power gain \\
        Power accepted by the network             & Operating power gain & \textbf{Unnamed power gain} (Friis equation) \\
        Power to load without network  & Insertion power gain  & Not used \\
    \bottomrule
    \end{tabular}
\end{table}

\section{Active Antenna Available Power}

\cl{
To generalize the Friis transmission formula to antenna system with active beamforming networks, we need a meaningful way to define the available power in \eqref{eq:friis} at the output of the receiving array. This can be done using active antenna available power \cite{IEEE_Std_145-2013}. 

The power at the output of an active array includes electronic gains and scaling coefficients in the beamforming network. The absolute level of the power is, in a sense, meaningless for an active array with electronic gains and beamformer weights. The net gain from an array element port to the analog to digital converters in a digitally beamformed system or MIMO array may be \mbox{40--50$\,$dB} or more in a practical system. Digital signal processing blocks often use shifting bit windows to keep computational results within the range of a field programmable gate array (FPGA) or another processor, making the overall scale factor in signal processing chains arbitrary. The scaling in beamforming weights is also arbitrary, as the antenna pattern and the signal-to-noise ratio (SNR) at the beamformer output are unaffected by any change in the overall scale factor.}

\cl{
For an active receiving array, ratios of output powers are more meaningful. In applications such as radiometry, we compare beam output power levels for different antenna-pointing directions to estimate the source flux density. The formed beam output power can be given a physical meaning with a comparison value or level calibration. What is needed is a way to scale the beamformer output so that it becomes physically meaningful.

To provide well-defined and measurable powers and figures of merit for active receivers, we use noise-based antenna parameter definitions and the concept of isotropic noise response \cite{warnick2010unified,warnick2012noise,IEEE_Std_145-2013}. 
For an active array, electronic gains and scaling in digital signal processing and beamformer coefficients can be determined using the response of the array to an isotropic external thermal noise environment, or the isotropic noise response $P_\text{iso}$. The isotropic noise response can be measured using the antenna Y factor method \cite{IEEE_Std_149-2021,warnick2022noise,buck2023experimental}. The isotropic noise response relative to the power $k_\mathrm{B} T_0 B$, where $k_\mathrm{B}$ is Boltzmann's constant, $T_0 = 290\,$K, and $B$ is the system noise equivalent bandwidth, is a measure of the electronic gains and beamformer coefficients in the receiver system \cite{IEEE_Std_145-2013}. Scaling the output power for an active array by $k_\mathrm{B} T_0 B/P_\text{iso}$ renders the active antenna effectively passive and gives the available power at the terminals of a passive antenna with the same radiation pattern as the active array, or the active antenna available power. 

The active antenna available power refers to the beam output power to the reference plane at the antenna terminals or ports. It has a number of practical applications, as it provides a way to define and measure effective area and aperture efficiency for active receiving arrays, and it can be directly compared to the system noise budget given in terms of antenna temperatures. We will use the active antenna available power to generalize the Friis transmission formula to active arrays in the near and far field cases. 
}

\section{Generalized Friis Transmission Formulas for the Ideal, Open-Circuit Loaded Case}
\label{sec:oc}

\begin{figure}
    \centering
	\includegraphics[width=0.95\linewidth]{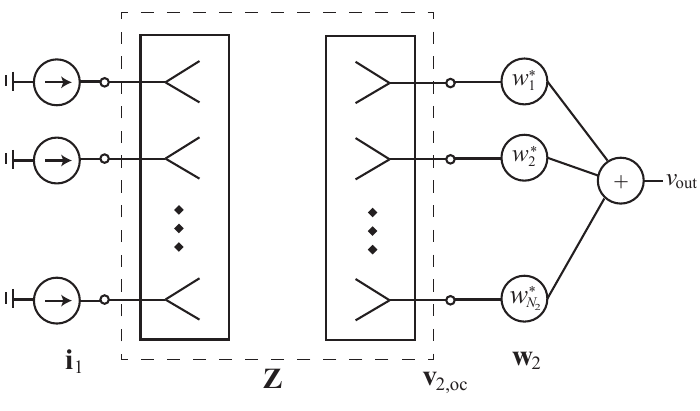}
	\caption{Transmitting array 1 fed by ideal current sources and receiving array 2 with ideal voltage sensors having infinite input impedance and a beamforming network (case A). For case B, the direction of transmission and reception is reversed. \cl{The dashed line enclosing both antennas represents a multiport network with impedance matrix $\Z$.}}
	\label{fig:system_beamformer_oc}
\end{figure}

Before considering the more complex case of transmitting and receiving arrays with arbitrary loads, we will use a simpler system with ideal sources and loads to illustrate the important concepts in the generalization of the Friis transmission formula to multiport communication systems. A transmitting array is excited by ideal current sources. At the receiver, ideal voltage sensors with infinite input impedance sample voltages from the array for beamforming to produce a scalar system output (Figure\ \ref{fig:system_beamformer_oc}). 

\cl{In case A, array 1 transmits and array 2 receives. As the active antenna available power requires a precise definition of the receiver output port, we explicitly show the receive array beamforming network. On the transmit side, only the amplitudes and phases of the currents at the array input ports matter and the network that realizes these excitations is unimportant and need not be shown explicitly. For case B, array 2 transmits and the beamforming network is moved to array 1.}

The arrays are modeled as an $(N_1 + N_2)$-port network with impedance matrix
\begin{equation}\label{eq:Z}
  \Z = \begin{bmatrix}
  \Z_{11} & \Z_{12} \\
  \Z_{21} & \Z_{22}
  \end{bmatrix}.
\end{equation}
Each port of the transmitting and receiving arrays are terminated with open-circuit loads\footnote{To maintain the open-circuit loading condition on the receiving array, we will form a beam at the receiving array using ideal zero admittance voltage sensors. The open-circuit loads eliminate the dependence of the impedances looking into array 1 on the loads terminating the array 2 element ports and leads to the simplest analysis in the impedance matrix formulation.}. A similar analysis could be made with short circuit loads and ideal current sensors using admittance matrices. We assume that the arrays and propagation medium are linear, time-invariant, and reciprocal. Since the system is reciprocal, the impedance matrix $\Z$ in \eqref{eq:Z} is symmetric, so that $\Z = \Z^\trans$. 

Until the progress made in \cite{ERPEE7,ERPEE8} for the case of network loads on the receive side instead of the beamformer considered here, symmetry of the bidirectional power transfer between arrays 1 and 2 has been an open problem. The goal is to study the reciprocity of the multiport antenna system in two cases: (A) array 1 transmitting with excitation currents $\iv_1$ and array 2 receiving with beamformer coefficients $\wv_2$; and (B) array 2 transmitting with currents $\iv_2$ and array 1 receiving with beamformer coefficients $\wv_1$. Since the system is reciprocal, there should be a relationship between the power transfer from array 1 to 2 and the reverse. 

In case A, array 1 is excited with ideal current sources with strengths given by the $N_1$-element column vector $\iv_1$. Array~2 receives an incident field from array~1 resulting in open-circuit voltages given by the $N_2$-element column vector 
\begin{equation}
    \vv_\text{2,oc} = \Z_{21} \iv_1.
\end{equation} 
Ideal voltage sensors with zero admittance followed by complex beamformer weights $\wv_2$ form a beam with voltage output 
\begin{equation}\label{eq:vout}
    v_\text{out} = \wv_2^\herm \vv_\text{2,oc} = \wv_2^\herm \Z_{21} \iv_1,
\end{equation}
where~$^\mathrm{H}$ denotes Hermitian conjugate and where by convention in array signal processing the receiver beamformer weights are applied with a complex conjugate so that the received power has the form of a matrix inner product. 

To generalize the Friis formula, we need the power accepted by array 1 and the output power at array 2 after beamforming. 
The power accepted by array 1 is
\begin{equation}
    P_\text{in,1} = \dfrac{1}{2} \iv_1^\herm \RE{\Z_{11}} \iv_1.
\end{equation}
where $\RE{\M}$ is the element-by-element real part of the matrix $\M$. 
This provides the denominator in the unnamed gain in \eqref{eq:friis}. The beam output power at array 2 is 
\begin{equation}\label{eq:Pout}
  P_\text{out} = \dfrac{1}{2 R} |v_\text{out}|^2,
\end{equation}
where the beam output voltage is given in \eqref{eq:vout}. Beams in a MIMO system or other array systems are usually formed in digital signal processing, and the absolute level of the output power has no meaning. As is customary in the array signal processing community, we will drop the constant in~\eqref{eq:Pout}, i.e., set $R = 1\,\Omega$.

\subsection{Active antenna available power}

The isotropic noise response of array 2 follows from Twiss's theorem \cite{noise_twiss_55,warnick2018phased}, and is
\begin{equation}
  P_\text{iso} = 8 k_\mathrm{B} T_0 B \wv_2^\herm \RE{\Z_{22}} \wv_2,
\end{equation}
where the units in this expression match the units of the beam output power in \eqref{eq:Pout}. 
Multiplying~\eqref{eq:Pout} by $k_\mathrm{B} T_0 B/P_\text{iso}$ and using~\eqref{eq:vout} gives the array~2 active antenna available power 
\begin{equation}\label{eq:P2a_oc}
    P_\mathrm{A,2} = \frac{|\wv_{2}^\herm \vv_\text{2,oc}|^2}{8 \, \wv_2^\herm \RE{\Z_{22}} \wv_2}
\end{equation}
at the formed beam output. This is the available power at the terminals of an equivalent passive, reciprocal antenna with the same radiation pattern as the beam formed at array 2 with weights $\wv_2$. Equation \eqref{eq:P2a_oc} provides the available power in the Friis transmission formula \eqref{eq:friis} for an active receiving array. 

\subsection{Active Antenna Unnamed Power Gain}

Combining the input power of array~1 with the active antenna available power at the output of beamformed array~2 gives the active antenna unnamed power gain for case~A, 
\begin{equation}\label{eq:gauoc}
    \GAUact = 
    \frac{P_\mathrm{A,2}}{P_\text{in,1}}
    =
    \frac{|\wv_{2}^\herm \Z_{21} \iv_\text{1A}|^2}{4 \, \iv_\text{1A}^\herm \RE{\Z_{11}}, \iv_\text{1A} \wv_2^\herm \RE{\Z_{22}} \wv_2}
\end{equation}
where \cl{the superscript $^\text{a}$ on $\GAUact$ denotes an active antenna parameter as defined in \cite{IEEE_Std_145-2013}. The subscript $_\text{A}$ on the current vector $\iv_\text{1A}$ indicates that array 1 is transmitting and array 2 is receiving. In case B, the direction of transmission is reversed.}

In case B, array 2 is excited as a transmitter with currents $\iv_\text{2B}$ and the received open-circuit voltages at array 1 are combined with beamformer coefficients $\wv_1$. The \cl{active antenna unnamed power gain}  is 
\begin{equation}\label{eq:gbuoc}
    \GBUact = 
    \frac{ |\wv_{1}^\herm \Z_{12} \iv_\text{2B}|^2}
    {4 \, \iv_\text{2B}^\herm {\rm Re}[ \Z_{22} ] \iv_\text{2B}  \wv_1^\herm {\rm Re}[\Z_{11}] \wv_1}.
\end{equation}
The expressions \eqref{eq:gauoc} and \eqref{eq:gbuoc} show that for an open-circuit loaded multiport antenna system in the near field and a linear, time-variant, and reciprocal but otherwise arbitrary propagation environment, the Friis transmission formula generalizes to an \cl{active antenna unnamed power gain}  for each link direction. 


\subsection{Link direction symmetry condition}

In general, $\GAUact$ and $\GBUact$ are different. But if we excite array~2 in such a way that its radiation pattern is the same as the receiving pattern in case~A, and if we choose beamformer coefficients for array~1 such that its receiving pattern in case~B is the same as its case~A radiation pattern, the \cl{active antenna unnamed power gains} for cases~A and~B can be shown to be equal.  

Array~2 radiates as a transmitter with the same radiation pattern as the receive beamformer when the transmit mode element input currents have the same magnitudes and phases as the receive mode voltage combining coefficients~$\wv_2^*$ \cite[(5.19)]{warnick2018phased}, with a complex conjugate to match the usual convention where beamformer weights are applied with a Hermitian conjugate as in \eqref{eq:vout}. Thus, array~2 transmits and receives with the same pattern when the array~2 beamformer weights in case~A are related to the currents at the array~2 ports in case~B according to 
\begin{equation}\label{eq:w2}
  \wv_2 \sim \iv_\text{2B}^* .
\end{equation}
Similarly, array~1 receives with the same pattern as when excited as a transmitter with
\begin{equation}\label{eq:w1}
  \wv_1 \sim \iv_\text{1A}^* .
\end{equation}
As mentioned above, the overall scale factor in the beamformer weights can be chosen arbitrarily, so we have used a proportionality relationship in these constraints on the beamformer weights. Because array 1 has the same radiation pattern on transmit (case A) as in receive (case B) and similarly for array 2, we expect $\GAUact = \GBUact$ when the above conditions hold. 

Using these conditions in the \cl{active antenna unnamed power gains} for transmission from array 1 to array 2 (case A) and from array 2 to array 1 (case B), we have
\begin{align}\label{eq:gauoc1}
  \GAUact &= \frac{\left|\iv_\text{2B}^\trans \Z_{21} \iv_\text{1A}\right|^2}
  {4 \, \iv_\text{1A}^\herm {\rm Re}[\Z_{11} ] \iv_\text{1A} \iv_\text{2B}^\herm {\rm Re}[\Z_{22}] \iv_\text{2B}}
  \\
  \GBUact &= 
    \frac{ \left|\iv_\text{1A}^\trans \Z_{12} \iv_\text{2B}\right|^2}
    {4 \, \iv_\text{2B}^\herm {\rm Re}[ \Z_{22} ] \iv_\text{2B}  \iv_\text{1A}^\herm {\rm Re}[\Z_{11}] \iv_\text{1A}}
    \label{eq:gbuoc1}
\end{align}
Because the system is reciprocal, $\Z_{21} = \Z_{12}^\trans$, the relationships in~\eqref{eq:gauoc1} and~\eqref{eq:gbuoc1} show that the symmetry condition \mbox{$\GAUact = \GBUact$} for the generalized Friis transmission formulas in \eqref{eq:gauoc1} and \eqref{eq:gbuoc1} hold when the beamformer weights in~\eqref{eq:w2} and~\eqref{eq:w1} are used. 

The generalized Friis transmission formulas in \eqref{eq:gauoc} and \eqref{eq:gbuoc} and the associated reciprocity condition hold for bidirectional transmission in a multiport antenna system, including the strongly coupled near-field case, but with idealized loads and voltage sensors. In the next section, we will generalize the Friis transmission formula to antenna arrays with arbitrary source and load networks. 

\section{Generalized Friis Transmission Formulas for Multiport Communication Systems}

\begin{figure}
	\centering
	\includegraphics[width=0.95\linewidth]{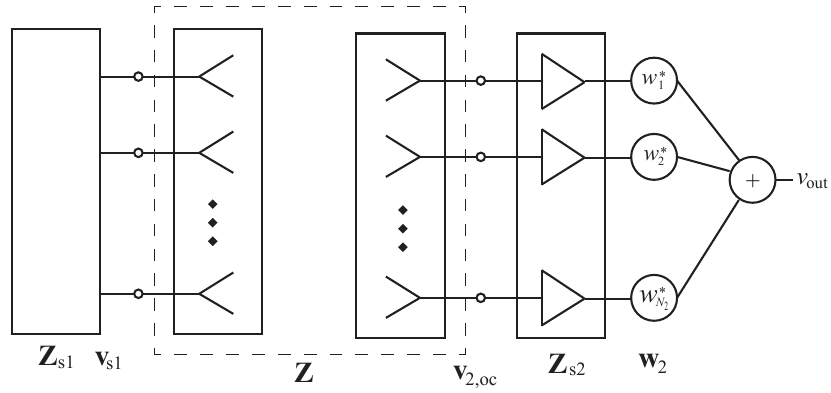}
	\caption{Arrays 1 and 2 with generic source and load networks and beamformer on array 2. This system diagram represents any array communication channel where signals are transmitted by array 1, received by array 2, amplified, and formed into an output that is a linear combination of the received signals, such as a point-to-point wireless link or an eigenchannel of a MIMO system.}
	\label{fig:system}
\end{figure}

We now turn to the case of a multiport antenna system with generic load networks, as shown in Figure~\ref{fig:system}. The arrays are terminated with source or load networks~$\Z_\mathrm{s1}$ and~$\Z_\mathrm{s2}$. In case~A, source~1 excites array~1 with open-circuit voltages~$\vv_\text{s1}$. Array~2 receives the incident field with open-circuit voltages~$\vv_\text{2,oc}$. While it may seem that this practical system with non-ideal loading networks is more complicated, we will show that its analysis is similar to the ideal open-circuit loaded case in Section~\ref{sec:oc} with the addition of coupling between array~2 with its loading network and the array~1 input impedance and vice versa. In the near field, the load on one array affects the input impedances of the other, and this must be accounted for in the case of generic source and load networks. 

The input impedances looking into the arrays are
\begin{align}\label{eq:zin1}
    \Z_\text{in,1} &= \Z_{11} - \Z_{12} (\Z_\text{s2} + \Z_{22})^{-1} \Z_{21},
   \\
    \Z_\text{in,2} &= \Z_{22} - \Z_{21} (\Z_\text{s1} + \Z_{11})^{-1} \Z_{12}.
\end{align}
The open-circuit voltages at the ports of array 2 induced by the radiated wave from array 1 are
\begin{align}
  \vv_\text{2,oc} &= \Z_{21} (\Z_\text{s1} + \Z_{11})^{-1}  \vv_\text{s1} \\
  &= \underbrace{\Z_{21} (\Z_\text{s1} + \Z_{11})^{-1} (\Z_\text{s1} + \Z_\text{in,1})}_{\M_\mathrm{A}} \iv_\text{1},
\end{align}
where $\Z_\text{s1}$ and $\Z_\text{s2}$ are the impedance matrices of the networks that load arrays 1 and 2, which we assume to be reciprocal. The loaded voltages at the ports of array 2
\begin{equation}
    \vv_2 = \Z_\text{s2} (\Z_\text{in,2} + \Z_\text{s2})^{-1} \vv_\text{2,oc}
\end{equation}
are amplified via receiver chains followed by analog or digital beamforming with complex weight vector $\wv_2$ to form the output voltage
\begin{equation}\label{eq:Q2def}
    v_\text{out} = \wv_2^\herm 
    \underbrace{g \, \Z_\text{s2} (\Z_\text{in,2} + \Z_\text{s2})^{-1}}_{\Q_2} 
    \vv_\text{2,oc},
\end{equation}
where $g$ is the voltage gain of the identical receiver chains. 

We define the equivalent open-circuit voltage beamformer weights
\begin{equation}
    \wv_\text{2,oc} = \Q_2^\herm \wv_\text{2}
\end{equation}
with which the beamformer output can be written in terms of the open-circuit voltages at array 2 as 
\begin{equation}
    v_\text{out} = \wv_\text{2,oc}^\herm \vv_\text{2,oc}.
\end{equation}
Generalizing \eqref{eq:P2a_oc} to the array system with generic source and load networks, the active antenna available power at array 2 becomes
\begin{equation}\label{eq:PA2}
    P_\mathrm{A,2} = \frac{|\wv_\text{2,oc}^\herm \vv_\text{2,oc}|^2}{8\, \wv_\text{2,oc}^\herm \RE{\Z_\text{in,2}} \wv_\text{2,oc}}.
\end{equation}
The power accepted by array 1 is
\begin{equation}
    P_\text{in,1} = \dfrac{1}{2} \RE{\iv_1^\herm \Z_\text{in,1} \iv_1},
\end{equation}
where 
\begin{equation}\label{eq:i1}
    \iv_1 = {(\Z_\text{in,1} + \Z_\text{s1})^{-1}} \vv_\text{s1}
\end{equation}
is a vector of currents flowing into array~1 expressed in terms of the source open-circuit voltages.

The \cl{active antenna unnamed power gain}  with array 1 transmitting and array~2 receiving is
\begin{equation}\label{eq:GAU}
    \GAUact = 
    \frac{|\wv_\text{2,oc}^\herm \vv_\text{2,oc} |^2}{4 \, 
    \wv_\text{2,oc}^\herm \RE{\Z_\text{in,2}} \wv_\text{2,oc} \iv_\text{1A}^\herm
    \RE{ \Z_\text{in,1} } \iv_\text{1A}
    }.
\end{equation}
For case~B, with array~2 transmitting and array~1 receiving, by the symmetry of the system, the \cl{active antenna unnamed power gain}  is 
\begin{equation}\label{eq:GBU}
    \GBUact = 
    \frac{|\wv_\text{1,oc}^\herm \vv_\text{1,oc} |^2}{4 \, 
    \wv_\text{1,oc}^\herm \RE{\Z_\text{in,1}} \wv_\text{1,oc} \iv_\text{2B}^\herm
    \RE{\Z_\text{in,2}} \iv_\text{2B}
    },
\end{equation}
where 
\begin{equation}
  \wv_\text{1,oc} =  \Q_1^\herm \wv_1.
\end{equation}
Here, $\wv_1$ is a vector of beamformer weights used to combine the loaded voltages at the array~1 ports on receive, and $\Q_1$ is defined analogously to~$\Q_2$ in~\eqref{eq:Q2def}. Equations~\eqref{eq:GAU} and~\eqref{eq:GBU} generalize the Friis transmission formula to multiport antenna systems in the near-field case and in a complex propagation environment. 

The \cl{active antenna unnamed power gain}  can be written as a combination of formal weighted inner products, 
\begin{equation}
    \label{eq:GAUaRQ}
    \GAUact = 
    \frac{|\langle \wv, \iv \rangle_{\M_\mathrm{A}} |^2}{4 
    \langle \wv,  \wv \rangle_{\RE{\Z_\text{in,2}}}
    \langle \iv, \iv \rangle_{\RE{\Z_\text{in,1}}}
    }.
\end{equation}
This is the generalized vector angle between the excitation currents and beamformer weights. If the beamformer weights on both arrays are normalized to unit input power, the \cl{active antenna unnamed power gain}  is $|\langle \wv, \iv \rangle_{\M_\mathrm{A}} |^2 / 16$. Expression~\eqref{eq:GAUaRQ} has the form of a generalized, bivariate Rayleigh quotient. Furthermore, in view of Eq.\ \eqref{eq:Z21} in Section \ref{sec:far_field}, this expression is closely related to the result in \cite[Eq.\ 3]{chu1965approximate}.

\subsection{Link direction symmetry condition for multiport antenna systems}

Even if the multiport antenna system and propagation environment are reciprocal, in general, $\GAUact \neq \GBUact$. In case~A, with array~1 transmitting and array~2 receiving, the excitations~$\iv_1$ at the transmitter and the beamformer weights~$\wv_2$ at the receiver are arbitrary, as are the excitations~$\iv_2$ and weights~$\wv_1$ in case B with array 2 transmitting and array 1 receiving. In case~A, the transmitter may aim a formed beam directly at array~2, for example, whereas in case~B array~2 may transmit a beam in a different direction, leaving array~1 in a null of the radiation pattern, so that $\GAUact$ is large and $\GBUact$ is zero. 

For a symmetry condition between~$\GAUact$ and~$\GBUact$ to hold, each array must transmit and receive with the same antenna pattern. This is accomplished using conditions similar to~\eqref{eq:w2} and \eqref{eq:w1}. In case~A, array~1 transmits with arbitrary excitations $\iv_\text{1A}$. To have the same pattern in case~B, array~1 must receive with beamformer weights on array~1 such that the complex conjugated open-circuit referenced beamformer combining coefficients are proportional to the transmit excitations from case~A, so that $\wv_\text{1,oc} \sim \iv_\text{1A}^*$ \cite{warnick2018phased}. Similarly, we require that the array~2 receiving pattern in case~A be the same as its radiation pattern in case~B, which obtain when $\wv_\text{2,oc} \sim \iv_\text{2B}^*$. It is worth noting that the array excitation currents in cases~A and~B are produced by the source excitation voltages~$\vv_\text{s1}$ and~$\vv_\text{s2}$, respectively, as given by \eqref{eq:i1} in case~A and its counterpart for case~B. 

Making these substitutions for the beamformer weights in \eqref{eq:GAU} and \eqref{eq:GBU} leads to
\begin{align}
    \GAUact &= 
    \frac{\left|\iv_\text{2B}^\trans \Z_{21} (\Z_\text{s1} + \Z_{11})^{-1} (\Z_\text{s1} + \Z_\text{in1}) \iv_\text{1A} \right|^2}{4 \,
    \iv_\text{2B}^\herm {\rm Re}[\Z_\text{in,2}] \iv_\text{2B}
    \iv_\text{1A}^\herm {\rm Re}[ \Z_\text{in,1} ] \iv_\text{1A}
    }
  \\
    \GBUact &= 
    \frac{\left|\iv_\text{1A}^\trans \Z_{12} (\Z_\text{s2} + \Z_{22})^{-1} (\Z_\text{s2} + \Z_\text{in2}) \iv_\text{2B} \right|^2}{4 \,
    \iv_\text{1A}^\herm {\rm Re}[\Z_\text{in,1}] \iv_\text{1A}
    \iv_\text{2B}^\herm {\rm Re}[ \Z_\text{in,2} ] \iv_\text{2B}
    }
\end{align}
The denominators of the two unnamed gains are equal. It remains to prove that the numerators are equal as well, and it suffices to show that $\M_\mathrm{A}^\trans = \M_\mathrm{B}$. Inserting \eqref{eq:zin1} and using the symmetry of $\Z$, we have 
\begin{align}
  \M_\mathrm{A}^\mathrm{T} &=  (\Z_\text{s1} + \Z_{11} - \Z_{12} (\Z_\text{s2} + \Z_{22})^{-1} \Z_{21}) (\Z_\text{s1} + \Z_{11})^{-1} \Z_{12} \nonumber \\
  &=  \Z_{12} - \Z_{12} (\Z_\text{s2} + \Z_{22})^{-1} \Z_{21} (\Z_\text{s1} + \Z_{11})^{-1} \Z_{12}
  \nonumber  \\
  &=  \Z_{12}(\Z_\text{s2} + \Z_{22})^{-1}\left[\Z_\text{s2} + \Z_{22} + \Z_{21} (\Z_\text{s1} + \Z_{11})^{-1} \Z_{12} \right]
  \nonumber  \\
  &= \Z_{12} (\Z_\text{s2} + \Z_{22})^{-1} (\Z_\text{s2} + \Z_\text{in2}) \nonumber  \\
  &= \M_\mathrm{B}
\end{align}
from which it follows that $\GAUact = \GBUact$ with beamformers \eqref{eq:w1} and \eqref{eq:w2} for the case of generic source and load networks.

Equations \eqref{eq:GAU} and \eqref{eq:GBU} and the reciprocity condition $\GAUact = \GBUact$ with $\wv_\text{1,oc} \sim \iv_\text{1A}^*$ and $\wv_\text{2,oc} \sim \iv_\text{2B}^*$ are the main results of this paper. These formulas generalize the Friis equation \eqref{eq:friis} to complex multiport antenna systems in arbitrary reciprocal propagation environments, including the case where one array is in the near field of the other. 
With a constraint on the beamformer coefficients to make the pattern of each array the same on transmitting and receiving, the \cl{active antenna unnamed power gains} become equal. This provides a generalized link direction symmetry condition for multiport antenna systems. 


\section{Embedded Element Patterns and Reduction to the Friis Equation}
\label{sec:far_field}


To provide a connection to the familiar far-field form of the Friis equation, we will show that the \cl{active antenna unnamed power gains} in the generalized Friis transmission formulas~\eqref{eq:GAU} and~\eqref{eq:GBU} factor, in the far-field limit, into the gains of antenna~1 and~2 and the inverse free space path loss. 
To accomplish this, we will use the antenna array embedded element pattern (EEP) formulation \cite{craeye2011review,davidson2022contemporary}. The EEP formulation provides far-field approximations for the impedance matrix transfer functions~$\Z_{21}$ and~$\Z_{12}$ in \eqref{eq:Z} in terms of array radiation field patterns that can be calculated or measured. 

EEPs are defined as the radiated fields of a multiport antenna system, with each port driven by a known excitation. In general, the EEPs depend on the loads that terminate undriven elements \cite{kelley2005embedded,warnick2020embedded}. The embedded element patterns for array~1 and~2 are denoted by $\Ev^1_m$, $m = \{ 1,2,\ldots,N_1 \}$ and $\Ev^2_n$, $n = \{ 1,2,\ldots,N_2 \}$, respectively. The EEPs for array 1 are 3D field vectors with components $\Ev^1_m = E^1_{m,x} \hat{\boldsymbol{x}} + E^1_{m,y} \hat{\boldsymbol{y}} + E^1_{m,z} \hat{\boldsymbol{z}}$ and similarly for array 2. The excitation is an ideal current source with unit strength $I_0 = 1\,$A, and the undriven elements are terminated with open-circuit loads. Even though we define or measure the EEPs with open-circuit loads, we can use network theory to find the fields radiated by the array with any combination of excitations and any load network on the element ports. 

In terms of the EEPs, the far-field approximation of the transfer function $\Z_{21}$ is given by 
\cite[Eq.\ 5.3]{warnick2018phased}
\begin{equation}\label{eq:Z21}
  Z_{21,mn} = \frac{2 \J \lambda r \T{e}^{\J k r} }{\eta |I_0|^2} \Ev^2_m(\rv_1) \cdot \Ev^1_n(\rv_2),
\end{equation}
where $\rv_2$ is the phase center of array 2 with respect to which the EEPs were measured, $\rv_1$ is the phase center of array 1, distance $r = |\rv_2 - \rv_1|$, wavenumber $k = 2 \pi/\lambda$, and $\eta$ is the impedance of free space. The expression \eqref{eq:Z21} is derived from the reciprocity principle and requires that the antenna arrays and the surrounding propagation medium be reciprocal. 

Using \eqref{eq:Z21} in \eqref{eq:GAU}, the \cl{active antenna unnamed power gain}  becomes
\begin{align}
    \GAUact &= 
    \frac{ \lambda^2 r^2|
    \Ev_1 \cdot \Ev_2 
    |^2}
    {\eta^2 \iv_\text{2}^\herm \RE{\Z_\text{in,2}} \iv_\text{2}
    \, \iv_1^\herm \RE{ \Z_\text{in,1} } \iv_1
    },
\end{align}
where the incident field at the location of array 2 radiated by array 1 with excitations $\iv_1$ is 
\begin{align}
    \Ev_1 &= \sum_{n = 1}^{N_1} i_{1,n} \Ev^1_n(\rv_2) 
\end{align}
and the field radiated by array 2 as a transmitter with excitations $\iv_2 = \wv_\text{oc,2}^*$ at the location of array 1 is 
\begin{align}
    \Ev_2 &= \sum_{m = 1}^{N_2} i_{2,m} \Ev^2_m(\rv_1).
\end{align}  
We are considering case A, with array 1 in transmit mode and array 2 receiving, but we have effectively used the reciprocity principle in \eqref{eq:Z21} to relate the voltages received by array 2 to the EEPs radiated by array 2 in transmit mode. This allows both arrays 1 and 2 to be characterized using radiated fields as given by the respective EEPs, whether the arrays are transmitting or receiving.

The polarization efficiency is
\begin{equation}
    \eta_\text{pol} = \frac{|\Ev_1 \cdot \Ev_2|^2}{|\Ev_1 |^2 |\Ev_2|^2}
\end{equation}
with which we can write
\begin{align}
    \GAUact &= 
    \frac{ \lambda^2 r^2 |\Ev_1|^2 |\Ev_1|^2 \eta_\text{pol}
    }
    {\eta^2 \iv_\text{2}^\herm \RE{\Z_\text{in,2}} \iv_\text{2}
    \, \iv_1^\herm \RE{ \Z_\text{in,1} } \iv_1
    }.
\end{align}
Recognizing that $P_\text{in,1} = \iv_1^\herm {\rm Re}[ \Z_\text{in,1} ] \iv_1/2$ is the power into array 1 with excitation currents $\iv_1$ and $P_\text{in,2} = \iv_\text{2}^\herm {\rm Re}[\Z_\text{in,2}] \iv_\text{2}/2$ is the power into array 2 with excitation currents $\iv_2$, the unnamed power gain becomes
\begin{align}
    \GAUact &= \lambda^2 r^2 
    \frac{ |\Ev_1|^2 }{2 \eta P_\text{in,1}}
    \frac{ |\Ev_2|^2 }{2 \eta  P_\text{in,2}}
    \eta_\text{pol}.
\end{align}
In the far-field approximation, the power density radiated by array 1 with excitation $\iv_1$ is $S_1 = |\Ev_1|^2 /(2 \eta)$, and similarly $S_2 = |\Ev_2|^2/(2 \eta)$. We then have
\begin{align}
    \GAUact &= 
    \frac{ 4 \pi r^2 S_1 }{P_\text{in,1}}
    \frac{ 4 \pi r^2 S_2 }{P_\text{in,2}}
    \eta_\text{pol}
    \left( \frac{\lambda}{4 \pi r} \right)^2
\end{align}
We recognize the first and second factors as the gains $G_1$ and $G_2$ of arrays 1 and 2, from which in the case of unit polarization efficiency \eqref{eq:friis} follows directly. Thus, in the far field, the \cl{active antenna unnamed power gain}  reduces to the familiar form of the Friis transmission equation and is independent of the loading networks $\Z_\text{s1}$ and $\Z_\text{s2}$, since, in the far-field limit and under realistic impedance conditions, $\mathbf{Z}_\text{in,1} \approx \mathbf{Z}_{11}$ and $\mathbf{Z}_\text{in,2} \approx \mathbf{Z}_{22}$.

For the reverse direction, $\GBUact$ can be similarly reduced to the Friis equation.
If~\eqref{eq:w1} is enforced, the gain~$G_1$ from case~A is equal to the gain of array~1 in case~B, and if~\eqref{eq:w2} is enforced, then~$G_2$ is the same in cases~A and~B. 

In the case of a non-free space propagation environment, the EEPs can be computed or measured in the embedded environment, and the generalized Friis formula holds with gains $G_1$ and $G_2$ computed from the EEPs in the propagation environment. In such a case, the far-field approximation used in~\eqref{eq:Z21} demands that transmitter, receiver and scattering environment are all in the far-field of each other.

\section{Connecting the Active Antenna Unnamed Power Gain With Earlier Results}  

For multiport antenna systems, the active antenna unnamed power gain or the power transfer from a transmitter to the received beam output depends on the excitations at the transmitting array and the beamformer coefficients at the receiving array. The dependence of the power transfer on the receiving array weights can be eliminated by considering the available power at the receiving array network ports (i.e., the power accepted by a conjugate matched load attached to the receiving array). In earlier work, the ratio of the available power at the receiving array network ports to the power accepted by the transmitting array has been defined as the unnamed power gain without the ``active antenna'' qualifier \cite{ERPEE7,ERPEE8}. The key difference between this paper and the results in earlier work is that in this paper the receive array has a beamformer that creates a linear combination of the signals at each array port, whereas in earlier work the output of the receive array is defined to be the available power at the ports of the receiving array network. 

\subsection{Unnamed Power Gain as a Maximum of Active Antenna Available Power}

The active antenna available power at the output of a beamformer is closely related to the available power from the array considered as a network. In essence, the maximum active antenna available power at the beamformer output over all sets of weighting coefficients is equal to the available power from the array network ports. Accordingly, we will show here that the active antenna unnamed power gain, when maximized over the receiving array beamformer coefficients, is equal to the unnamed power gain. In correspondence with the previous work~\cite[Section~XIII]{ERPEE7} and~\cite{ERPEE8}, we use $\GAU$ to denote the unnamed power gain between the ports of array~1 and the ports of array~2 in case~A, and $\GBU$ to denote the unnamed power gain between the ports of array~2 and the ports of array~1 in case~B. 
In this section, to increase the generality of the results, we do not assume that the system is reciprocal, so  the impedance matrix~$\Z$ is not necessarily symmetric. We use $H(\mathbf{M}) = (\mathbf{M}+\mathbf{M}^\herm)/2$ to denote the Hermitian part of a complex matrix~$\mathbf{M}$. For a reciprocal system, the impedance matrix is symmetric, in which case $H(\Z) = \RE{ \Z }$.

In case A, for a given excitation specified by the nonzero vector $\vv_\text{s1}$, it follows from \cite[Eq.~65]{ERPEE8} that $\GAU$ is given by 
\begin{equation}\label{equ. K-1 of draft}
\GAU= \dfrac
{\vv_\text{s1}^\herm
 \Z_\text{11s}^{-\herm}\,\mathbf{Z}_{21}^\herm
H(\Z_\text{in,2})^{-1} \,\mathbf{Z}_{21}
\Z_\text{11s}^{-1}\vv_\text{s1}}
{4 \,\vv_\text{s1}^\herm \Z_\text{in,1s}^{-\herm}
H(\Z_\text{in,1}) \,
\Z_\text{in,1s}^{-1}\vv_\text{s1}} \,.
\end{equation}
where $\Z_\text{11s} = \Z_\text{s1}+\mathbf{Z}_{11}$, $\Z_\text{in,1s} = \Z_\text{s1}+\Z_\text{in,1}$, and $\mathbf{A}^{-\herm} = \left(\mathbf{A}^{-1}\right)^\herm$. 
We have established in Section~III that $\GAUact$ is given by 
\begin{equation}\label{equ. K-2 of draft}
\GAUact= \dfrac
{ \left| \wv_\text{2,oc}^\herm 
\mathbf{Z}_{21} \Z_\text{11s}^{-1}
\vv_\text{s1} \right|^2}
{4 \, \wv_\text{2,oc}^\herm H(\Z_\text{in,2})
\wv_\text{2,oc} 
\vv_\text{s1}^\herm \Z_\text{in,1s}^{-\herm}
H(\Z_\text{in,1}) 
\Z_\text{in,1s}^{-1}\vv_\text{s1}} .
\end{equation}
Expanding the numerator leads to 
\begin{equation}\label{equ. K-3 of draft}
\GAUact= \dfrac
{ \wv_\text{2,oc}^\herm \,
\mathbf{Z}_{21} \Z_\text{11s}^{-1}
\vv_\text{s1} \vv_\text{s1}^\herm
\mathbf{Z}_\text{11s}^{-\herm} \mathbf{Z}_{21}^\herm
\, \wv_\text{2,oc} }
{4\,\wv_\text{2,oc}^\herm H(\Z_\text{in,2})
\wv_\text{2,oc} 
\vv_\text{s1}^\herm \Z_\text{in,1s}^{-\herm}
H(\Z_\text{in,1}) 
\Z_\text{in,1s}^{-1} \vv_\text{s1}} .
\end{equation}
Assuming that $H(\Z_\text{in,2})$ is positive definite, it follows from Theorem~12 of \cite{ERPEE7} that $\GAUact$ has a maximum value over all nonzero $\wv_\text{2,oc}$ in $\mathbb{C}^n$ given by 
\begin{equation}\label{equ. K-4 of draft}
\max_{\wv_\text{2,oc}} \GAUact = \dfrac
{\lambda_{\mathrm{max}}}
{4 \, \vv_\text{s1}^\herm \Z_\text{in,1s}^{-\herm}
H(\Z_\text{in,1}) \,
\Z_\text{in,1s}^{-1} \vv_\text{s1}},
\end{equation}
where $\lambda_{\mathrm{max}}$ is the largest eigenvalue of the matrix
\begin{align}\label{equ. K-5 of draft}
\mathbf{R}=
\mathbf{Z}_{21} \Z_\text{11s}^{-1}
\, \vv_\text{s1} \vv_\text{s1}^\herm \, 
\mathbf{Z}_\text{11s}^{-\herm}\,\mathbf{Z}_{21}^\herm H(\Z_\text{in,2})^{-1} .
\end{align}
Since $\vv_\text{s1}$ is a nonzero vector, $\vv_\text{s1} \vv_\text{s1}^\herm$ is of rank 1. Thus, $\mathrm{rank}\,\mathbf{R}\leqslant 1$. Assuming that $\mathbf{R}$ is not the null matrix, it follows from \cite[Section~1.4.P1]{Horn_3} that
\begin{align}\label{equ. K-6 of draft}
\lambda_{\mathrm{max}}=
 \vv_\text{s1}^\herm
\mathbf{Z}_\text{11s}^{-\herm} \, \mathbf{Z}_{21}^\herm
\,  H(\mathbf{Z}_\text{in,2})^{-1} 
 \mathbf{Z}_{21} \Z_\text{11s}^{-1}
\vv_\text{s1} \,.
\end{align}
%
%
Using \eqref{equ. K-6 of draft} in \eqref{equ. K-4 of draft}, we get 
\begin{equation}\label{equ. K-7 of draft}
\max_{\wv_\text{2,oc}} \GAUact = \dfrac
{\vv_\text{s1}^\herm
 \Z_\text{11s}^{-\herm}\,\mathbf{Z}_{21}^\herm
\,  H(\Z_\text{in,2})^{-1}
\mathbf{Z}_{21} \Z_\text{11s}^{-1}
\vv_\text{s1}}
{4 \, \vv_\text{s1}^\herm \Z_\text{in,1s}^{-\herm}
H(\Z_\text{in,1}) \,
\Z_\text{in,1s}^{-1}\vv_\text{s1}} \,.
\end{equation}
Comparing \eqref{equ. K-1 of draft} to \eqref{equ. K-7 of draft}, we obtain
\begin{equation}\label{equ. K-8 of draft}
\max_{\wv_\text{2,oc}} \GAUact =\GAU \,.
\end{equation}

We have proven that the set of the values of $\GAUact$ for all nonzero $\wv_\text{2,oc} \in \mathbb{C}^n$ has a greatest element which is equal to $\GAU$. We can of course also say that the set of the values of $\GBUact$ for all nonzero  $\wv_\text{1,oc} \in \mathbb{C}^m$ has a greatest element which is equal to $\GBU$. Accordingly, $G_{\mathrm{AU\,MAX}}$ and $G_{\mathrm{AU\,MIN}}$  defined in \cite[Section~XIII]{ERPEE7} and \cite{ERPEE8} are also related to $\GAUact$, and $G_{\mathrm{BU\,MAX}}$ and $G_{\mathrm{BU\,MIN}}$  defined in \cite[Section~XIII]{ERPEE7} and \cite{ERPEE8} are similarly related to $\GBUact$.

\subsection{Bounds and Reciprocity Relations for the \cl{Active Antenna Unnamed Power Gain} }

Based on the above proof, we have the following upper bounds on the active antenna unnamed power gains:
\begin{align}
  \GAUact(\iv_1,\wv_2)  \le G_{\mathrm{AU\,MAX}}, \label{eq:amax}  \\
  \GBUact(\iv_2,\wv_1)  \le G_{\mathrm{BU\,MAX}}, \label{eq:bmax} 
\end{align}
where the upper bounds are the maximum generalized eigenvalues defined in \cite{ERPEE7,ERPEE8}. Equality is achieved when the transmit excitations and receive beamformer weights are optimized for maximum power transfer. In these expressions, to help clarify the dependence of the gain quantities on the array excitations and weights, we have added the excitation and weight vectors explicitly as arguments. In the far field, the upper bounds in~\eqref{eq:amax} and~\eqref{eq:bmax} are reached when the transmitting array main beam (with maximum gain) is steered to the receiving array and the receiving array pattern main beam (with maximum gain) is steered to the transmitting array. 

If we optimize the receive beamformer weights for maximum power transfer, we have the lower and upper bounds 
\begin{align}  
  G_{\mathrm{AU\,MIN}} \le \max_{\wv_2} \GAUact(\iv_1,\wv_2) &\le G_{\mathrm{AU\,MAX}}, \\
  G_{\mathrm{AU\,MIN}} \le \max_{\wv_1} \GBUact(\iv_2,\wv_1) &\le G_{\mathrm{BU\,MAX}}. 
\end{align}
The remaining degree of freedom in~$\max_{\wv_2} \GAUact(\iv_1,\wv_2)$ is the transmitting array excitations~$\iv_1$. As the transmit excitations are adjusted, the transmitting array radiates more or less power towards the receiving array, which in the far field corresponds to the transmitting array radiation changing to put the receiving array in a pattern maximum (upper bound) or null (lower bound). The generalized eigenvalues from ~\cite{ERPEE7,ERPEE8} give the maximum and minimum values of the power transfer from the transmitting array to the receiving array. 

Finally, the reciprocity relationships associated with the active antenna unnamed power gain and the unnamed power gain are 
\begin{align}
  \GAUact(\iv_1,\iv_2^*) &=  \GBUact(\iv_2,\iv_1^*), \\
  G_{\mathrm{AU\,MIN}} &=  G_{\mathrm{BU\,MIN}}, \\
  G_{\mathrm{AU\,MAX}} &=  G_{\mathrm{BU\,MAX}}
\end{align}
between the active antenna unnamed power gain with beamformer on the receiving array and the maximum and minimum unnamed power gain \cite{ERPEE7,ERPEE8}.

\section{Numerical Results}

To illustrate the behavior of the active antenna unnamed power gain in the near and far fields, we give numerical results for a~$2 \times 2$ multiport antenna system. Arrays~1 and~2 each consist of two parallel half-wave dipoles spaced a half-wavelength apart. The arrays are oriented in parallel and facing each other with separation distance~$r$. The array loading networks are uncoupled with 50$\,\Omega$ self impedances. The impedance matrix for the system is modeled using a thin wire method of moments code. 

\begin{figure}
	\centering
    \includegraphics[width=\columnwidth]{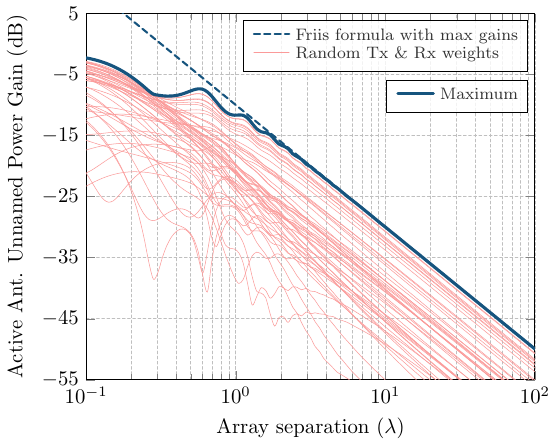}
	\caption{Active antenna unnamed power gains for a 2 $\times$ 2 multiport antenna system. In the far field, the maximum unnamed power gain converges to the Friis formula with maximum antenna gains combined with the best polarization efficiency. With random weights at both the transmitting and receiving arrays, the antenna gains of the arrays decrease and the \cl{active antenna unnamed power gain}  is lower than the maximum value of the unnamed power gain from~\cite{ERPEE8}. Notice the \cl{Friis} formula, the maximum gains, and polarization efficiency were evaluated in the far-field region and were then used for all array separations.}
	\label{fig:2x2_rand_tx_rx_weights}
\end{figure}

In Figure~\ref{fig:2x2_rand_tx_rx_weights}, the active antenna available gain with randomly chosen transmit excitation weights and receive beamformer weights is shown, along with the maximum value of the unnamed power gain from~\cite{ERPEE8}. With random weights, the active antenna unnamed power gain is lower than the maximum unnamed power gain from~\cite{ERPEE8}. Physically, in the far field limit where antenna gain is well defined, this is because the gains of arrays~1 and~2 with random weights are below their maximum values. As the array separation becomes large, the maximum unnamed power gain converges to the Friis formula~\eqref{eq:friis} with~$G_1$ and~$G_2$ computed with the array~1 and~2 beamformer weights optimized for maximum antenna gain. \cl{In the case of uniform arrays facing each other at broadside, the optimal weights in the far-field limit have equal amplitudes and phases, but in general this need not be the case.}

\begin{figure}
	\centering
    \includegraphics[width=\columnwidth]{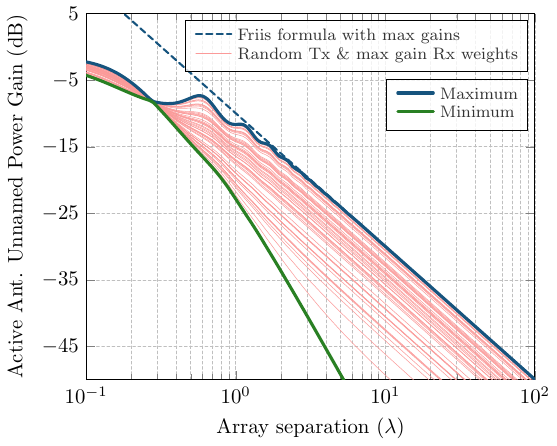}
	\caption{The active antenna unnamed power gain for a 2 $\times$ 2 multiport antenna system with receive beamformer weights optimized for maximum antenna gain and transmit beamformer weights chosen randomly lies between the maximum and minimum values from~\cite{ERPEE8}.}
	\label{fig:2x2_rand_tx_weights}
\end{figure}

Figure \ref{fig:2x2_rand_tx_weights} shows similar results but with maximum antenna gain beamformer weights at the receiver. The array~2 active antenna available power~\eqref{eq:PA2} is maximized when the beamformer weights are chosen according to
\begin{equation}\label{eq:w2max}
  \wv_\text{2,oc} = \RE{\Z_\text{in,2}}^{-1} \vv_\text{2,oc}
\end{equation}
With these beamformer weights, the active antenna available power is equal to the available power from the array~2 network ports (e.g., without the beamformer) \cite{warnick2018phased}
\begin{equation}
  P_\text{a,2} = \dfrac{1}{8} \vv_\text{2,oc}^\herm \RE{\Z_\text{in,2}}^{-1}\vv_\text{2,oc}
\end{equation}
In the far field with a plane wave incident field, \eqref{eq:w2max} is the maximum antenna gain beamformer solution~\cite{warnick2018phased}. The active antenna unnamed power gain with receive beamformer weights optimized for maximum antenna gain and transmit beamformer weights chosen randomly lies between the maximum and minimum values of the unnamed power gain defined using available power from the network as in~\cite{ERPEE8}. \cl{Interestingly, at one specific array separation, the maximum and minimum curves cross. At this separation, coupling from the transmitting array to the receiving array elements is not sensitive to the array excitations or beamformer weights.}

As a final step of this example, we can optimize the excitation weights at the transmitter for maximum power transfer. This can be done by using the generalized eigenvalue problem in~\cite{ERPEE8} to find the array~1 source excitations that maximize the unnamed power gain. If this is done, the active antenna available unnamed power gain with both transmit and receive weights optimized for maximum gain is equal to the upper bound in Figures~\ref{fig:2x2_rand_tx_rx_weights} and \ref{fig:2x2_rand_tx_weights}. \cl{The maximization is over all sets of beamformer weights, including highly oscillatory weights that could result in supergain.}


\begin{figure}
	\centering
	\includegraphics[width=0.9\linewidth]{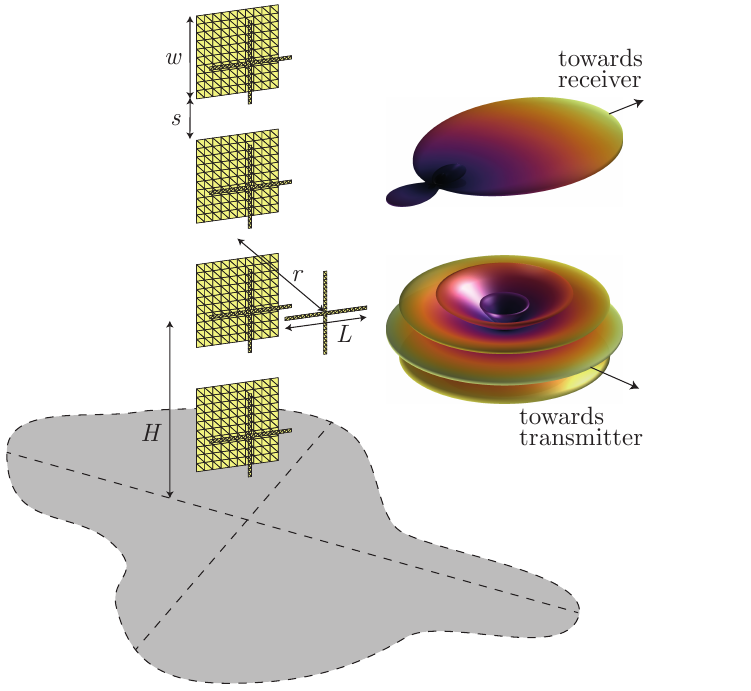}
	\caption{Model of a perfectly conducting transmitting dipole array consisting of four sections of crossed dipoles backed by a reflector. The receiving array is made of a single section of crossed dipoles. The system is fed and loaded by a diagonal matrix of~$50 \, \T{\Omega}$ loads. The system is positioned over \cl{an infinite perfectly conducting ground plane at height~$H$ equal to four wavelengths}. Length of the dipoles~$L \approx \lambda / 2$. The distance of the dipoles from the array reflector is~$2L/5$. The reflector is a square with side~$w = L$. Distance between the array sections is~$3L/2$. The figure also shows the directivity patterns of the transmitter (top) and receiver (bottom). The directivity patterns correspond to the maximum directivity in the direction of the transmitter/receiver. The directivity values are~$G_1 \approx 15 \, \T{dBi}$, $G_2 \approx 5.2 \, \T{dBi}$.}
	\label{fig:BTSsetup}
\end{figure}

A second numerical example shows the generality of the theory proposed in this paper. The example considers a link between a highly directive crossed dipole array and a simple crossed dipole receiver, see  Figure~\ref{fig:BTSsetup}. The antenna system is further positioned over a perfectly conducting ground plane to allow for indirect propagation paths. 

\begin{figure}
	\centering
	\includegraphics[width=\columnwidth]{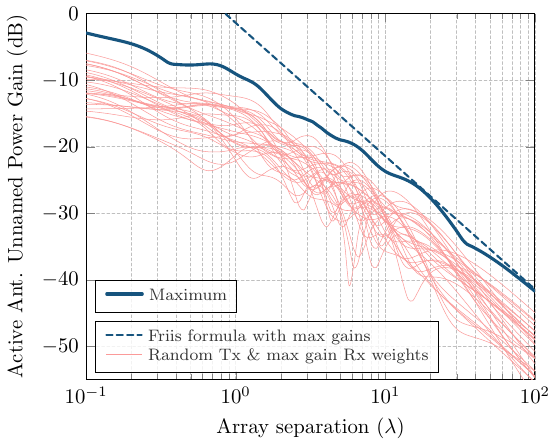}
	\caption{\cl{Active antenna unnamed power gain}  corresponding to the antenna system from Figure~\ref{fig:BTSsetup}. The minimum value of the unnamed power gain is not shown since its value is zero, the reason being an unequal number of transmitting and receiving elements~\cite{ERPEE8}.}
	\label{fig:Res1}
\end{figure}

The unnamed power gain~\eqref{equ. K-1 of draft} of this system is shown in Figure~\ref{fig:Res1}. The Friis transmission formula~\eqref{eq:friis} is also depicted in the figure and corresponds to the maximum value of the unnamed power gain in the far-field approximation. The use of crossed dipoles allows the radiation along the ground plane, which is used by the system to produce the highest unnamed gain. The directivity patterns corresponding to the maximum directivity of the transmitter and receiver along the ground plane are shown in Figure~\ref{fig:BTSsetup}. These directivities are employed by the Friis transmission formula~\eqref{eq:friis}. As shown in the body of this paper, the results for this example are invariant to numerical precision with respect to a change of transmitter and receiver in the \cl{maximum and minimum values of the active antenna unnamed power gain. Because a limited number of realizations for randomized values of the transmit and receive array are used, the curves do not reach the theoretical maximum value, but with an increased number of random weights, the maximum curve eventually will be reached.} 

\cl{In this last example it is worth noting that unlike in Fig.~\ref{fig:2x2_rand_tx_weights} the random realizations do not come close to the maximum and minimum values which demand specific excitations. For example, in this case, the minimum is equal to zero, which demands an excitation that eradicates any communication between the two arrays.} \clrev{As compared to Figs.\ \ref{fig:2x2_rand_tx_rx_weights} and \ref{fig:2x2_rand_tx_weights}, the configuration analyzed in Fig.\ \ref{fig:Res1} has a considerably larger solution space and the probability to approach the global optimum by the Monte Carlo procedure is lower. For this reason, the curves with random beamformer weights do not approach the upper bound as closely as in the simpler configurations.}

\section{Conclusion}

\cl{We have presented a generalized Friis transmission formula and derived a new property, the symmetry under link direction reversal for beamformed multiport antenna systems, which applies in the near field as well as the far field for arbitrarily complex reciprocal propagation environments.} The result is inspired by the generalizations of the Friis transmission formula proposed in \cite{ERPEE5,ERPEE7,ERPEE8}. The treatment connects noise-based active antenna terms and the isotropic noise response concept with the Friis transmission formula and highlights the importance of an unnamed power gain related to communication systems obtained from available output power divided by input power. We anticipate applications to MIMO communication systems, simultaneous transmit and receive systems, metrology for antenna arrays, and other applications of complex multiport antenna systems.





\begin{IEEEbiography}[{\includegraphics[width=1in,height=1.25in,clip,keepaspectratio]{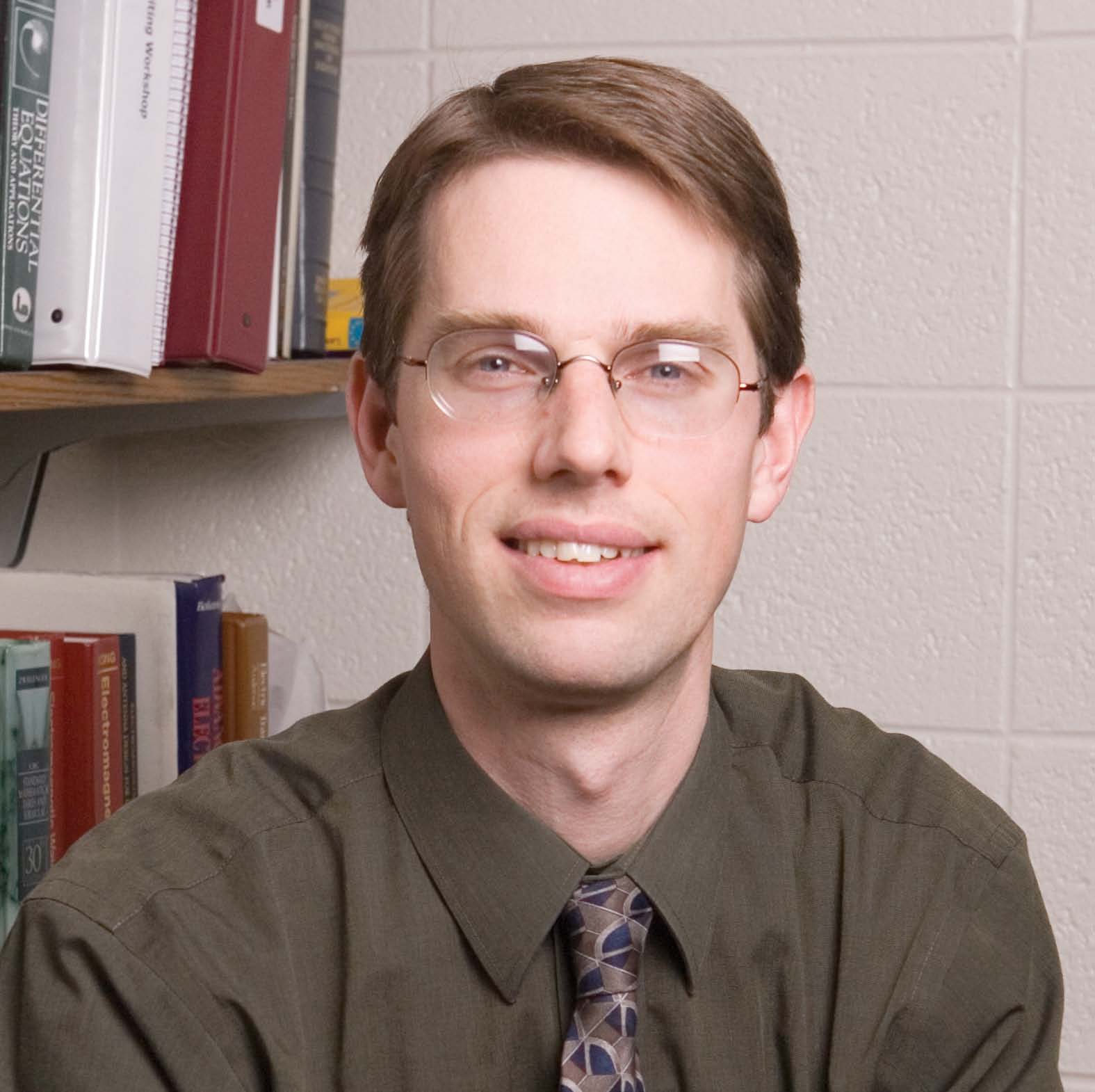}}] 
{Karl F. Warnick} (SM’04, F’13) received the BS degree in Electrical Engineering and Mathematics and the PhD degree in Electrical Engineering from Brigham Young University (BYU), Provo, UT, in 1994 and 1997, respectively. From 1998 to 2000, he was a Postdoctoral Research Associate and Visiting Assistant Professor in the Center for Computational Electromagnetics at the University of Illinois at Urbana-Champaign.  Since 2000, he has been a faculty member in the Department of Electrical and Computer Engineering at BYU, where he is currently a Professor. Dr.\ Warnick has published many books, scientific articles, and conference papers on electromagnetic theory, numerical methods, antenna applications, and high sensitivity phased arrays for satellite communications and radio astronomy. 
\end{IEEEbiography}

\begin{IEEEbiography}[{\includegraphics[width=1in,height=1.25in,clip,keepaspectratio]{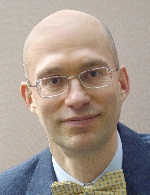}}] {Fr\'{e}d\'{e}ric Broyd\'{e}} (S'84 - M'86 - SM'01) was born in France in 1960. He received the M.S. degree in physics engineering from the Ecole Nationale Sup\'{e}rieure d'Ing\'{e}nieurs Electriciens de Grenoble (ENSIEG) and the Ph.D. in microwaves and microtechnologies from the Universit\'{e} des Sciences et Technologies de Lille (USTL).

He co-founded the Excem corporation in May 1988, a company providing engineering and research and development services. He is president of Excem since 1988. He is now also president and CTO of Eurexcem, a subsidiary of Excem. Most of his activity is allocated to engineering and research in electronics, radio, antennas, electromagnetic compatibility (EMC) and signal integrity.

Dr. Broyd\'{e} is author or co-author of about 100 technical papers, and inventor or co-inventor of about 90 patent families, for which 73 patents of France and 49 patents of the USA have been granted. He is a licensed radio amateur (F5OYE).
\end{IEEEbiography}

\begin{IEEEbiography}[{\includegraphics[width=1in,height=1.25in,clip,keepaspectratio]{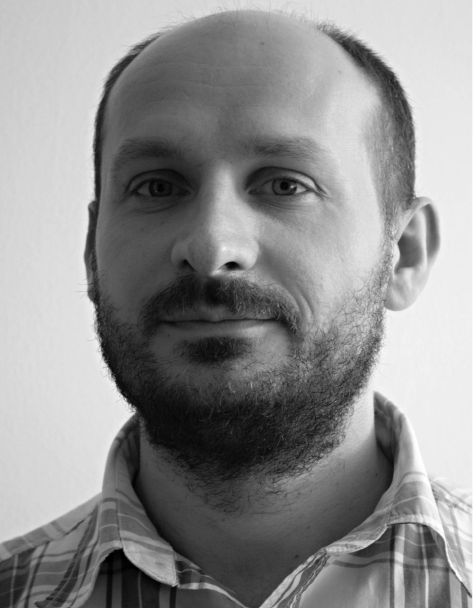}}]{Lukas Jelinek} was born in Czech Republic in 1980. He received his Ph.D. degree from the Czech Technical University in Prague, Czech Republic, in 2006. In 2015 he was appointed Associate Professor at the Department of Electromagnetic Field at the same university.

His research interests include wave propagation in complex media, electromagnetic field theory, metamaterials, numerical techniques, and optimization.
\end{IEEEbiography}

\begin{IEEEbiography}[{\includegraphics[width=1in,height=1.25in,clip,keepaspectratio]{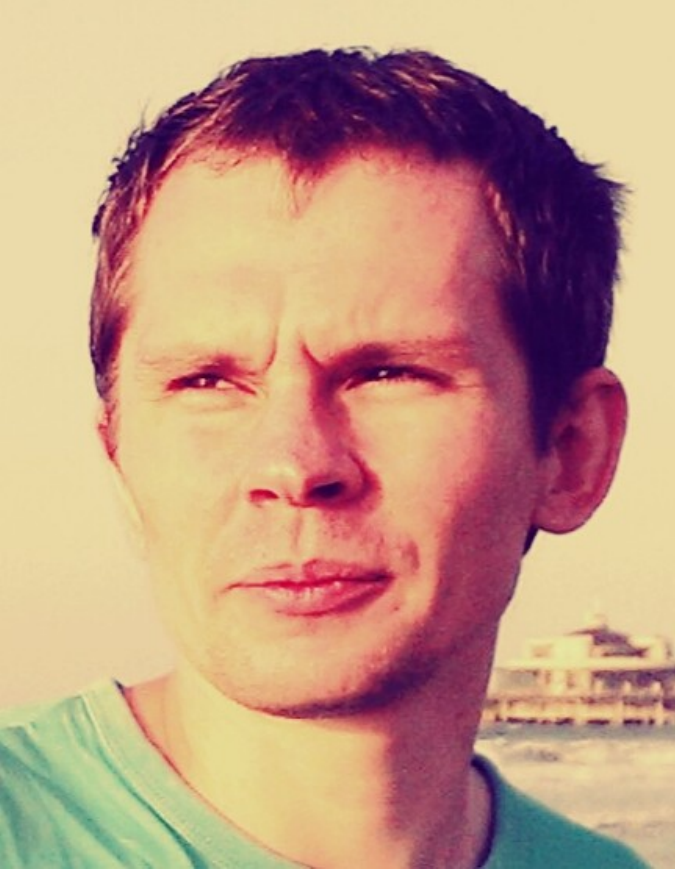}}]{Miloslav Capek}
(M'14, SM'17) received the M.Sc. degree in Electrical Engineering 2009, the Ph.D. degree in 2014, and was appointed a Full Professor in 2023, all from the Czech Technical University in Prague, Czech Republic.
	
He leads the development of the AToM (Antenna Toolbox for Matlab) package. His research interests include electromagnetic theory, electrically small antennas, antenna design, numerical techniques, and optimization. He authored or co-authored over 160~journal and conference papers.

Dr. Capek is the Associate Editor of IET Microwaves, Antennas \& Propagation. He was a regional delegate of EurAAP between 2015 and 2020 and an associate editor of Radioengineering between 2015 and 2018. He received the IEEE Antennas and Propagation Edward E. Altshuler Prize Paper Award~2023.
\end{IEEEbiography}

\begin{IEEEbiography}[{\includegraphics[width=1in,height=1.25in,clip,keepaspectratio]{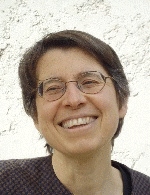}}] {Evelyne Clavelier} (S'84 - M'85 - SM'02) was born in France in 1961. She received the M.S. degree in physics engineering from the Ecole Nationale Sup\'{e}rieure d'Ing\'{e}nieurs Electriciens de Grenoble (ENSIEG).

She is co-founder of the Excem corporation, based in Maule, France, and she is currently CEO of Excem. She is also president of Tekcem, a company selling or licensing intellectual property rights to foster research. She is an active engineer and researcher.

Her current research areas are radio communications, antennas, matching networks, EMC and circuit theory.

Prior to starting Excem in 1988, she worked for Schneider Electrics (in Grenoble, France), STMicroelectronics (in Grenoble, France), and Signetics (in Mountain View, CA, USA).

Ms. Clavelier is the author or a co-author of about 90 technical papers. She is co-inventor of about 90 patent families. She is a licensed radio amateur (F1PHQ).
\end{IEEEbiography}

\end{document}